# Approximate analytical solutions for nonlinear Emden-Fowler type equations by differential transform method

Birol İbiş[*]

Department of Main Sciences, Turkish Air Force Academy, 34149 Yeşilyurt, İstanbul, Turkey

**Abstract** *In this paper, approximate analytical solutions of nonlinear Emden-Fowler type equations are obtained by the differential transform method (DTM). The DTM is a numerical as well as analytical method for solving integral equations, ordinary and partial diferential equations. To show the efficiency of the DTM, some examples are presented. Comparisons with exact solution show that the DTM is a powerful method for the solution of the nonlinear Emden-Fowler type equations.*

## 1. Introduction

The Emden-Fowler type of equations are singular initial value problems relating to second-order ordinary differential equations (ODEs) which have been used to model several phenomena in mathematical physics and astrophysics such as thermal explosions, stellar structure, the thermal behavior of a spherical cloud of gas, isothermal gas spheres, and thermionic currents[1-4]. Many problems in the literature of mathematical physics can be distinctively formulated as equations of Emden–Fowler type defined in the form

$$y'' + \frac{2}{x} y' + af(x)g(y) = 0, \quad y(0) = y_0, \; y'(0) = 0 \qquad (1)$$

where $f(x)$ and $g(y)$ are some given functions of $x$ and $y$ respectively. For $f(x) = 1$ and $g(y) = y^m$, Eq.(1) is the standard Lane-Emden equation that was used to model the thermal behavior of a spherical cloud of gas acting under the mutual attraction of its molecules[3] and subject to the classical laws of thermodynamics.

The Emden-Fowler type equations have significant applications in many fields of scientific and technical world, a variety of forms of $g(y)$ have been investigated by many researchers. Many methods including numerical and perturbation methods have been used to solve the Emden-Fowler type equations. The approximate solutions to the Emden-Fowler type equations were presented by Shawagfeh[5] and Wazwaz[6-8] using the Adomian decomposition method (ADM)[9]. Also Wazwaz applied ADM to solve the time-dependent Emden–Fowler type of equations[10]. Nouh, accelerated the convergence of power series solution of Lane-Emden type equations by using Euler-Abel transformation and Padé approximation[11]. Liao solved Lane-Emden type equations by applying homotopy analysis method (HAM)[13]. Also, Bataineh, Noorani and Hashim applied HAM to solve Emden-Fowler type of equations and the time-dependent Emden-Fowler type of equations[14,15]. In [17-19], the homotopy perturbation method (HPM)[20] is applied to Lane-Emden type equations. In [21-24], the variational iteration method (VIM)[25-28] is used to solve Emden-Fowler type of equations. Versy recently, Parand, Dehghan, Rezaei and Ghaderi applied Hermite function collocation (HFC) method[29].

The purpose of this paper is to introduce the differential transform technique as an alternative to existing methods in solving nonlinear Emden-Fowler type equations. The concept of differential transformation was first proposed by Zhou[9] in solving linear and nonlinear initial valued problems in electrical circuit analysis. The DTM is a numerical method based on the Taylor series expansion which constructs an analytical solution in the form of a polynomial. The traditional high order Taylor series method requires symbolic computation. However, the differential transform method obtains a polynomial series solution by means of an iterative procedure. With this method, the given differential equation and its related boundary conditions are transformed into a recurrence equation that finally leads to the solution of a system of algebraic equations as coefficients of a power series solution. This method is useful to obtain the exact and approximate solutions of linear and nonlinear differential equations. No need to linearization or discretization, large computational work. It has been used to solve effectively, easily and accurately a large class of linear and nonlinear problems with approximations.

---
[*] Corresponding author. Tel.: +90 212 6632490 (4335).
E-mail address: bibis@hho.edu.tr (B.İbiş).

## 2. Differential Transform Method (DTM)

Differential transform of $y(x)$ function is defined as follows:

$$Y(k) = \frac{1}{k!}\left[\frac{d^k y(x)}{dx^k}\right]_{x=0} \quad (2)$$

where $y(x)$ is the original function and $Y(k)$ is the transformed function. Differential inverse transform of $Y(k)$ is defined as:

$$y(x) = \sum_{k=0}^{\infty} Y(k) \cdot x^k \quad (3)$$

From Eqs. (2) and (3) we get

$$y(x) = \sum_{k=0}^{\infty} \frac{x^k}{k!}\left[\frac{d^k y(x)}{dx^k}\right]_{x=0} \quad (4)$$

which implies that the concept of differential transform is derived from Taylor series expansion, but the method does not evaluate the derivatives symbolically. However, relative derivatives are calculated by an iterative way which are described by the transformed equations of the original functions.

The following theorems that can be deduced from Eqs. (2) and (3) are given below:

**Theorem 1.** If $f(x) = \alpha g(x) \pm \beta h(x)$, then

$F(k) = \alpha G(k) \pm \beta H(k)$, where $\alpha$ and $\beta$ are constant.

**Theorem 2.** If $f(x) = \dfrac{d^n g(x)}{dx^n}$, then

$F(k) = (k+1)(k+2)\cdots(k+n)G(k+n)$

**Theorem 3.** If $f(x) = g(x) \cdot h(x)$, then

$$F(k) = \sum_{r=0}^{k} G(r) \cdot H(k-r)$$

**Theorem 4.** If $f(x) = g_1(x) \cdot g_2(x) \cdots g_n(x)$, then

$$F(k) = \sum_{k_{n-1}=0}^{k} \sum_{k_{n-2}=0}^{k_{n-1}} \cdots \sum_{k_1=0}^{k_2} G_1(k_1) \cdot G_2(k_2 - k_1) \cdots G_n(k - k_{n-1})$$

**Theorem 5.** If $f(x) = x^n$, then

$$F(k) = \delta(k-n) = \begin{cases} 1, & k = n \\ 0, & k \neq n \end{cases}, (n \text{ is a nonnegative integer})$$

## 3. The Differential Transformation of Nonlinear Functions

In this section, we will introduce a reliable and efficient algorithm to calculate the differential transform of nonlinear functions. As will be seen later, the proposed technique is based only on the differentiation and fundamental Theorems 1-5.

**Theorem 6.** If $f(y) = y^m$, then

$$F(k) = \begin{cases} Y(0)^m, & k = 0 \\ \dfrac{1}{Y(0)}\sum_{r=1}^{k}\dfrac{(m+1)r-k}{k}Y(r)F(k-r), & k \geq 1 \end{cases}$$

**Proof.** From the definition of transform,

$$F(0) = \left[y^m(x)\right]_{x=0} = y^m(0) = Y^m(0) \quad (5)$$

Now, taking a differentiation of $f(y) = y^m$ with respect to $x$, we get:

$$\frac{df(y)}{dx} = my^{m-1}(x)\frac{dy(x)}{dx} \quad (6)$$

or equivalently,

$$y(x)\frac{df(y(x))}{dx} = mf(y(x))\frac{dy(x)}{dx} \quad (7)$$

Application of the differential transform to Eq. (7) gives:

$$\sum_{r=0}^{k} Y(r)(k-r+1)F(k-r+1) = m\sum_{r=0}^{k}(r+1)Y(r+1)F(k-r) \quad (8)$$

From Eq.(8),

$$(k+1)Y(0)F(k+1) = m\sum_{r=0}^{k}(r+1)Y(r+1)F(k-r) \\ -\sum_{r=1}^{k}Y(r)(k-r+1)F(k-r+1) \quad (9)$$

or equivalently,

$$(k+1)Y(0)F(k+1) = m\sum_{r=1}^{k+1}rY(r)F(k-r+1) \\ -\sum_{r=1}^{k}Y(r)(k-r+1)F(k-r+1) \quad (10)$$

$$(k+1)Y(0)F(k+1) = \sum_{r=1}^{k+1}\{(m+1)r-k-1\}Y(r)F(k-r+1) \quad (11)$$

Replacing *k+1* by *k* yields:

$$kY(0)F(k) = \sum_{r=1}^{k}\{((m+1)r-k)Y(r)F(k-r)\} \quad (12)$$

From Eq.(12), we get:

$$F(k) = \frac{1}{Y(0)}\sum_{r=1}^{k}\left\{\left(\frac{(m+1)r-k}{k}\right)Y(r)F(k-r)\right\} \quad (13)$$

Combining Eqs. (5) and (13), we obtain the transformed function of $f(y) = y^m$ as:

$$F(k) = \begin{cases} Y(0)^m, & k=0 \\ \frac{1}{Y(0)}\sum_{r=1}^{k}\frac{(m+1)r-k}{k}Y(r)F(k-r), & k\geq 1 \end{cases} \quad (14)$$

This proof shows that the differential transform of other more complicate nonlinear function can be computed in a similar way. We gives some nonlinear functions and their transforms following theorems.

**Theorem 7.** If $f(y) = e^{ay}$, then

$$F(k) = \begin{cases} e^{aY(0)}, & k=0 \\ a\sum_{r=0}^{k-1}\frac{r+1}{k}Y(r+1)F(k-1-r), & k\geq 1 \end{cases}$$

**Theorem 8.** If $f(y) = \ln(ay+b),\ ay+b > 0$, then

$$F(k) = \begin{cases} \ln(\alpha Y(0)+\beta), & k=0 \\ \frac{\alpha}{\beta+\alpha Y(0)}Y(1), & k=1 \\ \frac{\alpha}{\beta+\alpha Y(0)}[Y(k)-\sum_{r=0}^{k-2}\frac{r+1}{k}F(r+1)Y(k-1-r)], & k\geq 2 \end{cases}$$

**Theorem 9.** If $f(y) = \sin(\alpha y)$ and $g(y) = \cos(\alpha y)$, then

$$F(k) = \begin{cases} \sin(\alpha Y(0)), & k=0 \\ \alpha\sum_{r=0}^{k-1}\frac{k-r}{k}G(r)Y(k-r), & k\geq 1 \end{cases}$$

and

$$G(k) = \begin{cases} \cos(\alpha Y(0)), & k=0 \\ -\alpha\sum_{r=0}^{k-1}\frac{k-r}{k}F(r)Y(k-r), & k\geq 1 \end{cases}$$

**Theorem 10.** If $f(y) = \sinh(\alpha y)$ and $g(y) = \cosh(\alpha y)$, then

$$F(k) = \begin{cases} \sinh(\alpha Y(0)), & k=0 \\ \alpha\sum_{r=0}^{k-1}\frac{k-r}{k}G(r)Y(k-r), & k\geq 1 \end{cases}$$

and

$$G(k) = \begin{cases} \cosh(\alpha Y(0)), & k=0 \\ \alpha\sum_{r=0}^{k-1}\frac{k-r}{k}F(r)Y(k-r), & k\geq 1 \end{cases}$$

## 4. Applications

In this section, we apply DTM to obtain approximate solutions of the Emden–Fowler type of equations.

**Example 1.** (*The standart Lane-Emden equation*)

For $a=1,\ f(x)=1,\ g(y)=y^m$ and $y_0=1$, Eq.(1) is the standart Lane-Emden equation of index $m$.

$$y'' + \frac{2}{x}y' + y^m = 0,\ (x>0,\ m\geq 0) \quad (15)$$

subject to the boundary conditions

$$y(0)=1,\ y'(0)=0 \quad (16)$$

Substituting $m=0,\ 1$ and 5 into Eq.(15) leads to the exact solution

$$y(x) = 1 - \frac{1}{3!}x^2 \quad (17)$$

$$y(x) = \frac{\sin x}{x} \quad (18)$$

$$y(x) = \left(1+\frac{x^2}{3}\right)^{-1/2} \quad (19)$$

respectively. For other values of $m$, there are not any analytic exact solutions but series solutions are obtainable.

Multiplying both sides Eq.(15) by $x$,

$$xy'' + 2y' + xy^m = 0 \quad (20)$$

By using the above theorems, we obtain following recurrence relation

$$Y(k+1) = -\frac{1}{(k+1)(k+2)} G(k-1) \tag{21}$$

where $G(k)$ is the differential transform of $g(y) = y^m(x)$ in Theorem 6.

From Eq.(2), the boundary conditions given in Eq.(16) can be transformed as

$$Y(0) = 1, \quad Y(1) = 0 \tag{22}$$

Substituting Eq.(22) into Eq.(21) and by straightforward iterative steps, we get the following (for k=0,1,2,…,n) values $Y(k)$

$$Y(0)=1,\ Y(1)=0,\ Y(2)=-\frac{1}{6},\ Y(3)=0,$$
$$Y(4)=\frac{m}{120},\ Y(5)=0,\ Y(6)=-\frac{m(8m-5)}{3\cdot 7!},\cdots \tag{23}$$

Substituting $m = 0$, into Eq.(23) leads to Eq.(17),

$$y(x) = \sum_{k=0}^{10} Y(k) \cdot x^k = 1 - \frac{1}{6}x^2 \tag{24}$$

Substituting $m = 1$ into Eq.(23), the following series solution is obtained.

$$y(x) = \sum_{k=0}^{n} Y(k) \cdot x^k = 1 - \frac{1}{6}x^2 + \frac{1}{120}x^4 - \frac{1}{5040}x^6$$
$$+ \frac{1}{362880}x^8 - \frac{1}{39916800}x^{10} + \cdots \tag{25}$$

which is Taylor series of Eq.(18).

However, substituting $m = 5$ into Eq.(23), we get following series solution.

$$y(x) = \sum_{k=0}^{n} Y(k) \cdot x^k = 1 - \frac{1}{6}x^2 + \frac{1}{24}x^4 - \frac{5}{432}x^6$$
$$+ \frac{35}{10368}x^8 - \frac{7}{6912}x^{10} + \cdots \tag{26}$$

which is Taylor series of Eq.(19).

**Example 2.** (*The isothermal gas spheres equation*)

For $a = 1$, $f(x) = 1$, $g(y) = e^y$, Eq.(1) is the isothermal gas spheres equation

$$y'' + \frac{2}{x} y' + e^y = 0, \ (x > 0) \tag{27}$$

subject to the boundary conditions

$$y(0) = 0, \ y'(0) = 0 \tag{28}$$

This model can be used to view the isothermal gas spheres, where the temperature remains constant. For a thorough discussion of the formulation of Eq.(27), see [4]. A series solution of Eq.(27) is obtained by Liao[11], Ramos[14] and Wazwaz[16] as

$$y(x) \cong -\frac{1}{6}x^2 + \frac{1}{120}x^4 - \frac{1}{1890}x^6 + \frac{61}{1632960}x^8$$
$$- \frac{4087}{1796256000}x^{10} \tag{29}$$

Multiplying both sides Eq.(27) by $x$ and applying the Theorems 2,3,5 and 7 we obtain following recurrence relation

$$Y(k+1) = -\frac{1}{(k+1)(k+2)} \sum_{r=0}^{k} \delta(r-1) G(k-r) \tag{30}$$

where $G(k)$ is the differential transform of $g(y) = e^y$.

From Eq.(2), the boundary conditions given in Eq. (28) can be transformed as

$$Y(0) = Y(1) = 0 \tag{31}$$

From theorem 7, the differential transform $G(k)$ in Eq.(30) is,

$$G(0) = e^{Y(0)} = 1 \tag{32}$$

$$G(k) = \sum_{r=0}^{k-1} \frac{r+1}{k} Y(r+1) G(k-1-r), \ k \geq 1 \tag{33}$$

Using Eqs. (30)-(33), for $n = 10$, the following series solution is obtained:

$$y(x) = \sum_{k=0}^{10} Y(k) \cdot x^k = -\frac{1}{6}x^2 + \frac{1}{120}x^4 - \frac{1}{1890}x^6$$
$$+ \frac{61}{1632960}x^8 - \frac{629}{224532000}x^{10} \tag{34}$$

The resulting graph of the isothermal gas spheres equation in comparison to DTM and those obtained by Wazwaz by using ADM[16] is shown in Fig.1.

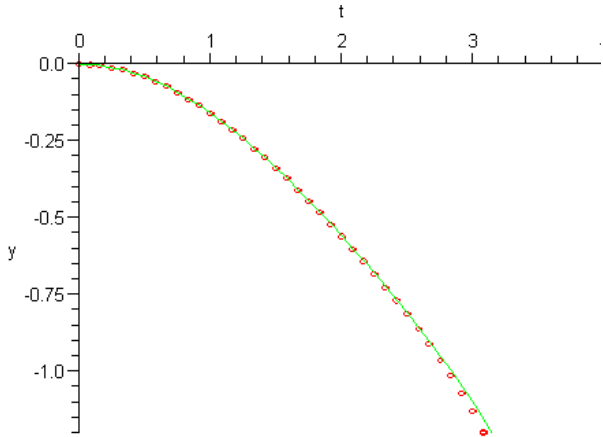

**Fig.1.** *Graph of isothermal gas sphere equation in comparison with Wazwaz solution [16].*

**Example 3.** For $a=1$, $f(x)=1$, $g(y)=\sinh(y)$, Eq.(1) is one of the Emden-Fowler type equations

$$y''+\frac{2}{x}y'+\sinh(y)=0, \quad (x>0) \tag{35}$$

subject to the boundary conditions

$$y(0)=1, \quad y'(0)=0 \tag{36}$$

Series solution of Eq.(35) is obtained by Wazwaz [6] as

$$y(x)\cong 1-\frac{e^2+1}{12e}x^2+\frac{e^4-1}{480e^2}x^4-\frac{2e^6+3e^2+3e^4+2}{30240e^3}x^6 \\ +\frac{61e^8+104e^6-104e^2-61}{26127360e^4}x^8 \tag{37}$$

Multiplying both sides Eq.(35) by $x$ and applying the Theorems 2,3,5 and 10 we obtain following recurrence relation

$$Y(k+1)=-\frac{1}{(k+1)(k+2)}\sum_{r=0}^{k}\delta(r-1)S(k-r) \tag{38}$$

where $S(k)$ is the differential transform of $g(y)=\sinh(y)$. From Eq.(2), the boundary conditions given in Eq. (36) can be transformed as

$$Y(0)=1, \quad Y(1)=0 \tag{39}$$

From Theorem 10, the differential transform $S(k)$ in Eq.(38) is,

$$S(0)=\sinh(Y(0))=\tfrac{1}{2}(e^1-e^{-1}) \\ S(k)=\sum_{r=0}^{k-1}\frac{k-r}{k}C(r)Y(k-r), \quad k\geq 1 \tag{40}$$

$$C(0)=\cosh(Y(0))=\tfrac{1}{2}(e^1+e^{-1}) \\ C(k)=\sum_{r=0}^{k-1}\frac{k-r}{k}S(r)Y(k-r), \quad k\geq 1 \tag{41}$$

where $C(k)$ is the differential transform of $f(y)=\cosh(y)$.

From Eqs. (38)-(41), for $n=10$, the following series solution is obtained:

$$y(x)\cong 1-\frac{e^2+1}{12e}x^2+\frac{e^4-1}{480e^2}x^4-\frac{2e^6+3e^2+3e^4+2}{30240e^3}x^6 \\ +\frac{61e^8+104e^6-104e^2-61}{26127360e^4}x^8 \\ -\frac{629e^{10}-1319e^8+902e^6-902e^4+1319e^2-629}{7185024000e^5}x^{10} \tag{42}$$

Eq.(42) is same Wazwaz's solution by using ADM[6].

**Example 4.** For $a=1$, $f(x)=1$, $g(y)=\sin(y)$, Eq.(1) is one of the Emden-Fowler type equations

$$y''+\frac{2}{x}y'+\sin(y)=0, \quad (x>0) \tag{43}$$

subject to the boundary conditions

$$y(0)=1, \quad y'(0)=0 \tag{44}$$

Series solution of Eq.(43) is obtained by Wazwaz[6] by using ADM is:

$$y(x)\cong 1-\frac{k_1}{6}x^2+\frac{k_1 k_2}{120}x^4+k_1\left(\frac{k_1^2}{3024}-\frac{k_2^2}{5040}\right)x^6 \\ +k_1 k_2\left(-\frac{113k_1^2}{3265920}+\frac{k_2^2}{362880}\right)x^8 \\ +k_1\left(\frac{1781k_1^2 k_2^2}{8988128000}-\frac{k_2^4}{39916800}-\frac{19k_1^4}{23950080}\right)x^{10} \tag{45}$$

where $k_1=\sin(1)$, $k_2=\cos(1)$.

Multiplying both sides Eq.(43) by $x$ and applying the Theorems 2,3,5 and 9 we obtain following recurrence relation

$$Y(k+1)=-\frac{1}{(k+1)(k+2)}\sum_{r=0}^{k}\delta(r-1)S(k-r) \tag{46}$$

where $S(k)$ is the differential transform of $g(y)=\sin(y)$. From Eq.(2), the boundary conditions given in Eq. (44) can be transformed as

$$Y(0)=1, \quad Y(1)=0 \tag{47}$$

From Theorem 9, the differential transform $S(k)$ in Eq.(46) is,

$$S(0) = \sin(Y(0)) = \sin(1)$$
$$S(k) = \sum_{r=0}^{k-1} \frac{k-r}{k} C(r) Y(k-r), \ k \geq 1 \quad (48)$$

$$C(0) = \cosh(Y(0)) = \cos(1)$$
$$C(k) = \sum_{r=0}^{k-1} \frac{k-r}{k} S(r) Y(k-r), \ k \geq 1 \quad (49)$$

where $C(k)$ is the differential transform of $\cos(y)$.

From Eqs.(46)-(49), for $n=10$, the following series solution is obtained:

$$y(x) \cong 1 - \frac{k_1}{6}x^2 + \frac{k_1 k_2}{120}x^4 + k_1 \left(\frac{k_1^2}{3024} - \frac{k_2^2}{5040}\right)x^6$$
$$+ k_1 k_2 \left(-\frac{113 k_1^2}{3265920} + \frac{k_2^2}{362880}\right)x^8 \quad (50)$$
$$+ k_1 \left(\frac{1781 k_1^2 k_2^2}{8988128000} - \frac{k_2^4}{39916800} - \frac{19 k_1^4}{23950080}\right)x^{10}$$

Eq.(50) is same Wazwaz's solution by using ADM[6].

**Example 5.** We consider the following nonlinear Emden-Fowler type equation[16].

$$y'' + \frac{5}{x}y' + 8a(e^y + 2e^{y/2}) = 0, \ (x>0) \quad (51)$$

subject to the boundary conditions

$$y(0) = y'(0) = 0 \quad (52)$$

Eq.(51) has the following analytical solution:

$$y(x) = -2\ln(1 + ax^2) \quad (53)$$

Multiplying both sides Eq.(51) by $x$ and applying the Theorems 2,3,5 and 8 we obtain following recurrence relation

$$Y(k+1) = \frac{-8a}{(k+1)(k+5)} \sum_{r=0}^{k} \delta(r-1)[E_1(k-r) + 2E_2(k-r)] \quad (54)$$

where $E_1(k)$ and $E_2(k)$ is the differential transform of $e^y$ and $e^{y/2}$ respectively.

From Eq.(2), the boundary conditions given in Eq. (52) can be transformed as

$$Y(0) = Y(1) = 0 \quad (55)$$

From Theorem 8, the differential transform $E_1(k)$ and $E_2(k)$ are,

$$E_1(0) = e^{Y(0)} = 1$$
$$E_1(k) = \sum_{r=0}^{k-1} \frac{r+1}{k} Y(r+1) E_1(k-1-r), \ k \geq 1 \quad (56)$$

$$E_2(0) = e^{\frac{1}{2}Y(0)} = 1$$
$$E_2(k) = \frac{1}{2}\sum_{r=0}^{k-1} \frac{r+1}{k} Y(r+1) E_2(k-1-r), \ k \geq 1 \quad (57)$$

From Eqs.(54)-(57), for $n=10$, the following series solution is obtained:

$$y(x) = \sum_{k=0}^{15} Y(k) \cdot x^k = -2ax^2 + a^2 x^4 - \frac{2}{3}a^3 x^6 + \frac{1}{2}a^4 x^8$$
$$- \frac{2}{5}a^5 x^{10} + \frac{1}{3}a^6 x^{12} - \frac{2}{7}a^7 x^{14} \quad (58)$$

which is Taylor series of Eq.(53).

**Example 6.** We finally consider the following nonlinear Emden-Fowler type equation[16]

$$y'' + \frac{8}{x}y' + 18ay = -4ay \ln y, \ (x>0) \quad (59)$$

subject to the boundary conditions

$$y(0) = 1, \ y'(0) = 0 \quad (60)$$

Eq.(62) has the following analytical solution:

$$y(x) = e^{-ax^2} \quad (61)$$

Multiplying both sides Eq.(59) by $x$ and applying the Theorems 2,3,5 and 7 we obtain following recurrence relation

$$Y(k+1) = \frac{-a}{(k+1)(k+8)} \left\{ \begin{array}{l} 18\sum_{r=0}^{k} \delta(r-1) Y(k-r) \\ +4\sum_{r=0}^{k}\sum_{m=0}^{r} \delta(m-1) Y(r-m) F(k-r) \end{array} \right\} \quad (62)$$

where $F(k)$ is the differential transform of $f(y) = \ln y$.

From Eq.(2), the boundary conditions given in Eq. (60) can be transformed as

$$Y(0) = 1, \ Y(1) = 0 \quad (63)$$

From Theorem 7, the differential transform $F(k)$ is,

$$F(0) = \ln(Y(0)) = 0$$
$$F(1) = \frac{1}{Y(0)} Y(1) = 0 \quad (64)$$
$$F(k) = \frac{1}{Y(0)}[Y(k) - \sum_{r=0}^{k-2} \frac{r+1}{k} F(r+1) Y(k-1-r)], \ k \geq 2$$

From Eqs.(62)-(64), for $n=10$, the following series solution is obtained:

$$y(x) = \sum_{k=0}^{15} Y(k) \cdot x^k = 1 - ax^2 + \frac{1}{2}a^2 x^4 - \frac{1}{6}a^3 x^6 + \frac{1}{24}a^4 x^8 \\ - \frac{1}{120}a^5 x^{10} + \frac{1}{720}a^6 x^{12} - \frac{1}{5040}a^7 x^{14}$$ (65)

which is Taylor series of Eq.(61).

## 5. Conclusion

In this study, Differential transform method is successfully applied to obtain numerical and/or analytical solutions of nonlinear singular initial value problems of Emden-Fowler Type equations. A symbolic calculation software package, MAPLE is used for all calculations The work emphasized our belief that the method is a reliable technique to handle these types of problems. Also, it provides the solutions in terms of convergent series with easily computable components in a direct way without using linearization, discretization or restrictive assumptions. The DTM offers great advantages of straightforward applicability, computational efficiency and high accuracy.